\newcommand{\He}{He~II $\lambda$4686}
\newcommand{\Halfa}{H${\alpha}$}

\newcommand{\Halp}{H${\alpha}$}

\newcommand{\Msol}{M$_{\odot}$}

\documentclass{iaus}
\usepackage{graphicx}

\title[ULX bubbles and counterparts] 
{Ultraluminous X-ray Sources: Bubbles and Optical Counterparts}

\author[short author list]   
{Manfred W. Pakull, 
 Fabien Gris\'e  \and Christian Motch}
\affiliation{Observatoire Astronomique de Strasbourg, France
 \break email: pakull@astro.u-strasbg.fr\\[\affilskip]}

\pubyear{2005}
\volume{230}  
\pagerange{}
\date{?? and in revised form ??}
\jname{Populations of High Energy Source in Galaxies}
\editors{Evert J.A. Meurs \& Guiseppina Fabbiano, eds.}
\begin{document}

\maketitle

\begin{abstract}
Optical studies of ultraluminous X-ray sources (ULX) in nearby galaxies 
have turned out to be instrumental in discriminating between various 
models including the much advertised intermediate mass black hole
hypothesis and various beaming scenarios. Here we report on 
ESO VLT and SUBARU observations of ULX that have revealed the parent stellar 
clusters with ages of some 60 million years in two cases. 
Thus we are able to derive upper limits of about 8 \Msol\ for the 
mass donors in these systems. The optical counterparts are dominated 
by X-ray heated accretion disks, and the discovery of the \He\  
emission line now allows to derive dynamical masses in these systems. 
Apparent radial velocity variations of 300 km/s have been detected in 
NGC 1313 X-2 which, if confirmed by further observations, would exclude 
the presence of IMBH in these systems. 
  
\end{abstract}

\firstsection 
\section{The enigma of ultraluminous X-ray sources}

One of the most significant results from recent X-ray studies 
of nearby galaxies is the discovery of a number of non-nuclear point 
sources (Ultraluminous X-ray sources - ULX) with apparent (isotropic) 
X-ray luminosities of 10$^{39}$--10$^{41}$ erg/s, a factor of 
10--1000 times brighter than typical luminous X-ray binaries in our 
Galaxy. 

From X-ray timing and spectral studies it is clear that ULXs are 
accreting compact objects. Therefore, their luminosity should not 
exceed the Eddington limit 
$L_{\rm E}=1.3\times10^{38}$~M/M$_{\odot}$~erg/s. 
One possible explanation is that the accreting object powering an ULX
is an intermediate-mass black hole (IMBH), with M $\sim$ 10$^2$-10$^4$ M$_{\odot}$
(\cite[Colbert \& Mushotzky 1999]{comu99}, 
\cite[Makishima \etal\ 2000]{maki00}, \cite[Miller, Fabian \& Miller 2004]{mfm04}). 
An alternative explanation is that 
the emission is beamed along the observer's line of sight, either  
by geometrical effects (\cite[King \etal\ 2001]{king01}), or due to 
relativistic jets like in microquasars (\cite[Fabrika \& Mesheryakov 2001]{fame01}), 
so that the true total luminosity does not exceed the 
Eddington limit of a stellar-mass black hole. Yet another 
possibility is that the accretor is a stellar BH 
genuinely emitting above the classical Eddington limit 
(\cite[Begelman 2002]{beg02}).

At the same time, one has to explain the high accretion rate (up to 
$\sim$ 10$^{-6}$ M$_{\odot}$/yr) required by the inferred 
luminosities. Wind accretion (typical for high-mass X-ray binaries) 
or Bondi-Hoyle type accretion from the interstellar medium are clearly too 
inefficient. It is more likely that the mass transfer occurs via 
Roche-lobe overflow from a donor star to the black hole. Stable, high mass transfer
can proceed on a nuclear time scale ($\leq$ several 10$^7$ yr), if   
the mass donor is massive - but not much more massive than the black hole
(cf. \cite[Rappaport, Podsiadlowski \& Pfahl 2005]{rap05}). 

\section{Bubble Nebul\ae}

An important piece of information for our understanding of ULX 
comes from the discovery of huge ionised bubble nebul\ae\ around a 
significant fraction of unobscured ULX in nearby galaxies 
(Pakull \& Mirioni 2002, 2003). It is even likely that all ULX sources 
are surrounded by such structures. Close to the X-ray source, we sometimes
observe X-ray ionisation effects which in the case of Holmberg~II X-1
has allowed to independently measure the (ionising) X-ray luminosity
(\cite[Pakull \& Mirioni 2002]{pami02}).  
The presence in the outer regions of strong [S~II] and [O~I] emission 
lines and of supersonic expansion speeds of 80-250 km/s derived from 
the width of \Halfa\ emission show that (at least the outer parts of) 
the bubbles are shock-excited rather then photoionised.

Among the best-studied large ULX bubbles are MH~9-11 around Holmberg~IX X-1 
(cf. Gris\'e \etal, this symposium) and NGC~1313 X-2 which both have 
diameters of about 500 pc, i.e. they are much larger than 
supernova remnants.  

One of the many interesting aspects of ULX bubbles is that we might estimate the 
kinetic energy involved in the ULX phenomenon. Two possibilities for bubble formation
have been discussed: 
either they were formed in an explosive event with kinetic energy $E_0$ (possibly the 
SN explosion that created the compact component in the ULX), or they are being 
inflated by ULX stellar wind/jet activity with mechanical energy $L_w$. 
Assuming for simplicity that energy is largely conserved 
(Sedov-Taylor solution), we have for the SNR case 
\begin{equation}
E_0 \approx 1.9\ 10^{52}\ erg\ R_2^3\ v_2^2\ n \qquad t_6 \approx 0.4\ R_2\ v_2^{-1}
 \label{Sedov}
\end{equation}
and for the wind/jet case
\begin{equation}
L_w \approx 3.8\ 10^{39}\ erg\, s^{-1}\ R_2^2\ v_2^3\ n \qquad t_6 \approx 0.6\ R_2\ v_2^{-1}
 \label{Wind}
\end{equation}
Here R$_2$ is the radius in units of 100 pc, and v$_2$ is the 
expansion velocity 
in units of 100 km/s of a bubble having an age of 10$^6$ t$_6$ yrs.  
The interstellar particle density, n, into which the bubble expands 
can be estimated from comparing the observed \Halp\ emission with the intensity 
I$_{\alpha} =$ 10$^{-6}$ I$_{\alpha,-6}$\ erg\ cm$^{-2}$ s$^{-1}$ sr$^{-1}$ of a fully 
radiative shock (\cite[Dopita \& Sutherland 1996]{dosu96}):
\begin{equation}
n \approx 0.6\ cm^{-3}\ I_{\alpha,-6}\ v_2^{-2.4 }
\label{density}
\end{equation}

Straightforward application of these equations to ULX bubbles yields 
typical ages of 10$^6$ yrs and densities in the range of 0.1--1.0 cm$^{-3}$. 
This results in very large energy/power requirements of $\sim$ 10$^{52-53}$ erg 
for the SNR case,  and $\sim$ 10$^{39-40}$ erg s$^{-1}$ for continuous inflation. 
However, a SNR is likely to expand into a region of very 
low density that has previously been excavated by the stellar winds from 
an evolving cluster (see below) before hitting the walls of that 
'superbubble'. In this way, energy requirements could well 
be ten times lower, and more akin to typical SN energies of 
10$^{51}$ erg. Another complication arises from the clumpiness of
the interstellar medium which results in smaller shock velocities in the
dense optically emitting clouds as compared to a higher velocity of the 
main shock in the intercloud medium. Taking into account this effect, 
\cite{blair81} estimated the kinetic of SNR to be: 
\begin{equation}
E_0 \approx 4\ 10^{50}\ erg\ R_2^3\ n(SII)
 \label{super}
\end{equation}
where n(SII) is the density in the recombination zone behind the shock, as measured
by the well-known forbidden [SII] line ratio. However, this method is not 
without severe problems either, as the energy calculated in this way appears to be  
positively correlated with the remnant diameter, an effect that may be related to the
magnetic pressure in the dense clouds.   

As outlined below, we favour the possibility of ongoing 
inflation by stellar winds/jets which would require for reasonable
mass loss rates $\leq$10$^{-6}$ \Msol/yr mildly relativistic ejection 
velocities akin to the famous v=0.26\,c jets in SS~433. However, inflation
into a cloudy medium would also lower the power requirements somewhat.

\section{Clusters}

Most massive stars are born in clusters or associations. 
Searching for such birthplaces of ULX thus opens the possibility 
to learn more about their ages
and possible formation scenarios. 
For the relatively isolated ULX in Holmberg~IX and in NGC~1313 X-2 we 
have been able to identify small faint clusters of stars that 
in each case are clearly associated with the ULX (cf. 
 Gris\'e \etal, this symposium).
\begin{figure}[!h]
  \begin{center}
    \begin{tabular}{cc}
      \resizebox{!}{6cm}{\includegraphics{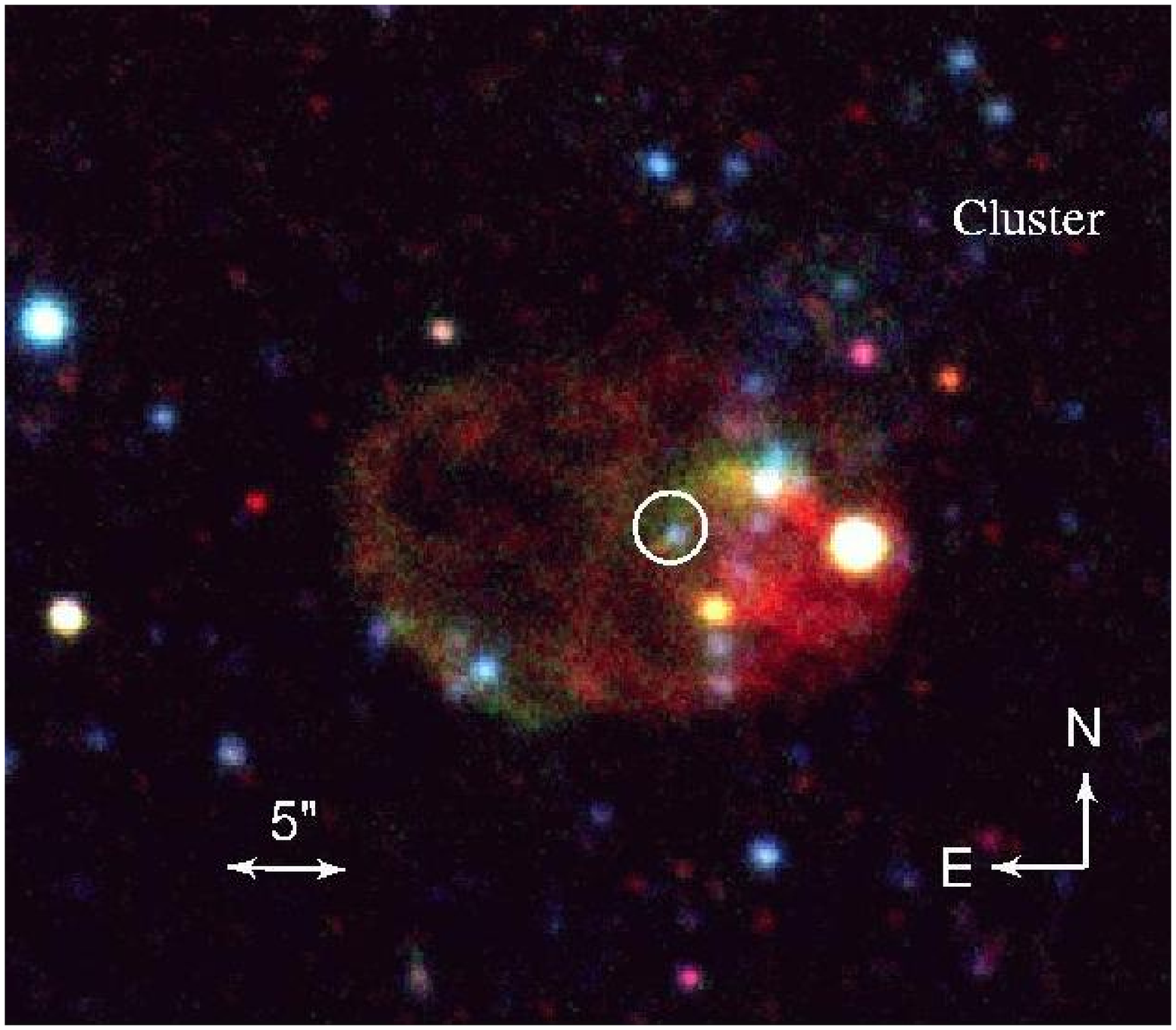}}&
      \resizebox{!}{6cm}{\includegraphics{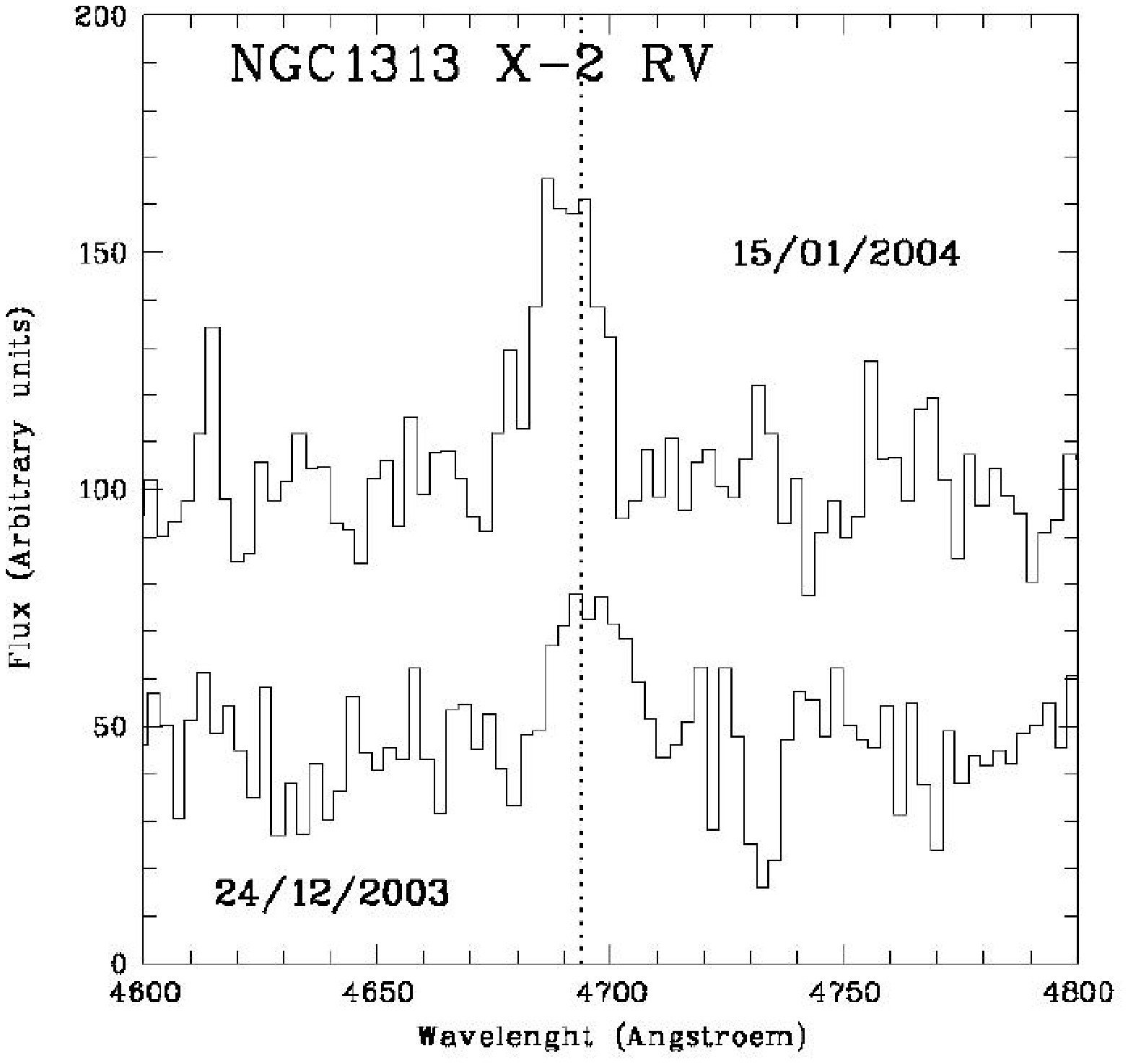}} \\
    \end{tabular}
    \caption[]{\\\hspace{\linewidth}
{\em LHS:}  Multicolor image (including \Halfa) of the 400 pc diameter bubble 
around the ULX NGC 1313 X-2. The Chandra error 
circle includes a close pair of stars (star C in \cite[Zampieri \etal\ 2004]{zam04}). 
The N-W component is the optical counterpart.\\
{\em RHS:} Blue range spectra of the counterpart taken 
3 weeks apart. Most prominent is the \He\ emission line which appears to have moved 
by 300 km/s.}
    \label{figures}
  \end{center}
\end{figure}
Multicolour 
photometry and isochrone fitting yield ages of some 40-70 
Myrs and total cluster masses of some 10$^3$ \Msol. Thus, 
all ionising O stars have already exploded a long time ago, implying 
again that the nebul\ae\ cannot possibly be photoionized by stellar 
EUV continua. This has an important consequence for the mass donor 
component in the ULX: it cannot be more massive then about 
8 \Msol. Furthermore, if the bubble nebul\ae\ indeed represent the 
remnants of the formation process of the ULX black holes, then
these explosions (which took place about 1 Myr ago) have taken place 
in an advanced stage of cluster evolution, i.e. the progenitor stars of 
the ULX accretors would not have been much more massive then the current donors 
(i.e. $\leq$ 10 \Msol). Such stars are of course not likely to become 
massive black holes. 

We are therefore left with the hypothesis that the ULX nebulae
represent 
\cite{Beg80} 'beambags' of jet inflation like the 'ears' in the (radio) nebula 
W~50 around SS~433.

\section{Optical counterparts}
 The optical counterparts of ULXs 
Holmberg~IX and NGC~1313 X-2 (see Fig. 1) have visual magnitudes 
of 22.9 and 23.4, respectively, and blue optical colors \mbox{(B-V$\sim$0.0)}; 
at a distance of 3--4 Mpc this translates into \mbox{M$_v$$\sim$\,-5}. Optical 
spectra taken with the ESO VLT and with the SUBARU telescope reveal the presence of 
stellar \He\ emission with equivalent width of 10 and 18 \AA, respectively. 
This high excitation line can be considered a hallmark of X-ray binaries, 
being formed in the X-ray heated disk around the compact component. 
In Galactic high mass X-ray binaries (such as the L$_X$=10$^{38}$ erg/s systems
Cen X-3, SMC X-1, LMC X-4, etc)  the corresponding EWs are more than an order
of magnitude smaller, attesting to their much smaller X-ray luminosities. 
Therefore, the presence of strong \He\ emission in ULX might also be taken 
as strong evidence against the X-ray beaming scenarios mentioned earlier.  
We note by the way that a \He\ emitting Wolf-Rayet star interpretation can 
be ruled out given the advanced age of the cluster. 

Strong support for an accretion disk 
interpretation comes also from its position 
in the famous \cite[van Paradijs \& McClintock (1994)] {vPM94}
$\Sigma$ -- M$_v$ diagram of low mass X-ray binaries (i.e. of X-ray 
ionised disks), at the very high-luminosity end of this relation. Here, 
$\Sigma$ is proportional to the optical light expected from an accretion disk 
that is heated by a given X-ray luminosity; note that we have assumed a semi-detached
binary of some 20 \Msol\ total mass and an orbital period of a few days.

The RHS of Fig. 1 illustrates two observations of the NGC~1313 X-2 counterpart
separated by about three weeks. It appears that the \He\ emission has varied in 
radial velocity (RV) by about $\sim$ 300 km/s around the velocity of the local 
H~I gas in the galaxy (dotted line). {\em If}\, this variation reflects real RV changes 
(as opposed to possible profile changes of the line having intrinsic FWHM $\sim$600 km/s),
then the object in the center of the 
accretion disk cannot be very massive, i.e. one could rule out the presence of an IMBH.

We finally mention variations by $\sim$0.2 mag amplitude over 9 nights 
in our B band photometry of the optical counterpart of NGC~1313 X-2. However, no stricly 
periodic signal
is seen, such as expected from ellipsoidal variations, and we ascribe these changes
to variable X-ray heating of the accretion disk.

\begin{acknowledgments}
We acknowledge collaboration with T.G. Tsuru, K. Sekigushi, A. Tajitsu and I. Smith.  
\end{acknowledgments}

\end{document}